\documentclass{article}
\usepackage{amssymb}

\usepackage{amsmath}

\input{tcilatex}

\begin{document}

\title{A Pragmatic Interpretation of Quantum Logic}
\author{Claudio Garola \\
Department of Mathematics and Physics, University of Salento \\
73100 Lecce, Italy\\
E-mail: Garola@le.infn.it}
\maketitle

\begin{abstract}
Scholars have wondered for a long time whether the language of quantum
mechanics introduces a quantum notion of truth which is formalized \ by
quantum logic (QL) and is incompatible with the classical (Tarskian) notion.
We show that QL can be interpreted as a pragmatic language $\mathcal{L}%
_{Q}^{P}$ of \textit{assertive} formulas, which formalize statements about
physical systems that are empirically \textit{justified} or \textit{%
unjustified} in the framework of quantum mechanics. According to this
interpretation, QL formalizes properties of the metalinguistic notion of 
\textit{empirical justification within\ quantum mechanics} rather than
properties of a quantum notion of truth. This conclusion agrees with a
general integrationist perspective that interprets nonstandard logics as
theories of metalinguistic notions different from truth, thus avoiding
incompatibility with classical notions and preserving the globality of
logic. By the way, some elucidations of the standard notion of quantum truth
are also obtained.\medskip

\end{abstract}

\section{\protect\bigskip Introduction}

Several years ago a \textit{formalized pragmatic calculus} $\mathcal{L}^{P}$
was constructed, based on some deep ideas by Dalla Pozza [Dalla Pozza and
Garola 1995]. This calculus extends a classical propositional calculus
stemming from Frege's \textit{ideographic language,} in which the \textit{%
assertion sign} is introduced as a constitutive part of the formulas of the
logical calculus [Frege, 1879, 1891, 1893, 1918; Reichenbach, 1947; Stenius,
1969]. $\mathcal{L}^{P}$ is obtained indeed by considering a classical
propositional calculus (CPC) with standard connectives and formation rules
for \textit{radical formulas}, and adding, besides the assertion sign, 
\textit{pragmatic connectives} and formation rules for \textit{assertive
formulas}. The radical formulas of $\mathcal{L}^{P}$ are then supplied with
a classical semantic interpretation, while the assertive\ formulas are
supplied\ with a pragmatic interpretation in terms of the notion of\ \textit{%
justification} (or \textit{proof}).

The first aim of this construction is providing a general framework in which
the conflict between the classical and the verificationist theories of truth
and meaning can be settled by integrating their perspectives. The
verificationist theories of truth, indeed, have been criticized by many
scholars, mainly because they assume that a proposition is true if and only
if asserting it is justified, which leads to identify the notions of truth
and justification [Russel, 1940, 1950;\ Carnap, 1949; Popper, 1969; Haack,
1978]. There are strong intuitive arguments that support the need of
avoiding this identification. In fact, the pragmatic notion of justification
presupposes the semantic notion of truth, for a proof of a sentence consists
in showing that the truth value of the sentence is \textit{true}. Moreover,
there are factual and logico-mathematical sentences that are \textit{%
undecidable}, i.e., epistemically inaccessible, though they have a truth
value [Carnap, 1932; Russell, 1940]. A sharp distinction between
justification and truth is therefore introduced in $\mathcal{L}^{P}$ via the
assertion sign, which has a purely pragmatic role and cannot be identified
with an alethic modality operator (it formalizes indeed the metalinguistic
notion of proof, which cannot be reduced to any acceptable notion of truth,
in the object language). Correspondingly, there are \ deep syntactic
differences between the part of $\mathcal{L}^{P}$ formalizing the properties
of the classical notion of truth (radical formulas) and the part of $%
\mathcal{L}^{P}$ formalizing the general properties of the notion of
justification (assertive formulas).

The second aim of the construction of $\mathcal{L}^{P}$ is\ showing that the
integration of the notions of truth and justification realized in $\mathcal{L%
}^{P}$ allows one to settle, in particular, the conflict between classical
and intuitionistic logic in a unified perspective. To this end two partially
overlapping structures, ACPC and AIPC (where A stands for \textit{assertive}%
) are singled out in $\mathcal{L}^{P}$\ that are isomorphic to CPC and to an
intuitionistic propositional calculus (IPC), respectively. The pragmatic
interpretation of $\mathcal{L}^{P}$ induces, through the isomorphism of AIPC
and IPC, an interpretation of IPC which recovers in a natural way the
standard Brower-Heyting-Kolmogorov (BHK) interpretation of this calculus in
terms of logical proof [Troelstra and Van Dalen, 1988]. The construction of $%
\mathcal{L}^{P}$ can thus help enlightening ``the mysteries of the
intuitionistic truth'' [Van Dalen, 1986].

The general perspective underlying the construction of $\mathcal{L}^{P}$ is
resumed in the Introduction of the paper by Dalla Pozza and Garola as
follows.

\begin{quote}
``The purpose of our interpretation is mainly philosophical. Indeed we aim
to settle the conflicts between classical and intuitionistic logic, and
between the classical (correspondence) and the intuitionistic
(verificationist) conceptions of truth and meaning (see Dummett, 1977, 1978,
1979, 1980; Prawitz, 1977, 1980, 1987); this will be done by introducing an 
\textit{integrated} perspective which preserves both the \textit{globality}
of logic (in the sense of the\ \textit{global pluralism}, which admits the
existence of a plurality of mutually compatible logical systems, but not of\
systems which are mutually incompatible or rivals, see Haack 1978, Chapter
12) and the \textit{classical} notion of \textit{truth as correspondence},
which we may consider \textit{explicated} rigorously by Tarski's semantic
theory (see Tarski 1933, 1944).''
\end{quote}

The ideas underlying the construction of $\mathcal{L}^{P}$ have started a
lively debate on pragmatics and related topics. Based on Dalla Pozza's
distinction between descriptive and expressive notions of norms, Bellin and
Dalla Pozza [2002], Bellin and Ranalter [2003] and Ranalter [2006] have
developed a pragmatic theory of obligations, assertions and causal
implication; White [2008] has presented a formal theory of actions; Carrara
and Chiffi [2013] have applied the logic for pragmatics framework to
knowability paradoxes. Moreover intuitionistic dualities have been explored
from the viewpoint of the logic for pragmatics, where co-intuitionism has
been regarded as a logic of hypotheses in relation with the intuitionistic
logic of assertions [Bellin and Biasi 2004; Biasi and Aschieri 2008; Bellin,
2014; Bellin 2015; Bellin et al., 2015a; Bellin et al., 2015b].

The quotation above from the paper by Dalla Pozza and Garola also reminds us
that there are several research fields in which non-classical notions of
truth are introduced, raising conflicts with classical logic (CL) similar to
the conflict between intuitionistic logic and CL. A typical example is
provided by quantum logic (QL), with its non-classical structure that is
claimed to imply a highly\ problematical notion of \textit{quantum truth} (a
huge literature exists on this topic; the interested reader can find a
general review of the attempts at constructing a logic for the language of
quantum mechanics till the early seventies in the classical book by Jammer
[1974], and more updated treatments and bibliographies in the books by Redei
[1998] and Dalla Chiara \textit{et al.} [2004]). One can then wonder whether
some of the foregoing conflicts\ can be settled by resorting to $\mathcal{L}%
^{P}$\ and embedding into it the non-classical structures that are
considered, as in the case of intuitionistic logic. The answer is positive,
as such an embedding has been realized for Girard's \textit{linear logic}
[Girard, 1987] by Bellin and Dalla Pozza [2002]. We aim to show in this
paper that\ a similar result can be achieved if QL is considered. The\
general scheme provided by $\mathcal{L}^{P}$ applies indeed not only when
the notion of proof is specified to be a \textit{logical proof}, as in the
case of IPC, but also if it is specified to be an \textit{empirical proof}
(or \textit{verification}, or \textit{empirical justification}) in the
framework of a specific theory.

Let us resume the main lines of our work.

By considering the notion of truth in quantum mechanics, we observe in Sect.
2 that our program has to face a deep problem from the very beginning.
Indeed, truth and verification are strictly entangled in the standard
interpretations of quantum mechanics. To avoid this problem we introduce a
generalization of $\mathcal{L}^{P}$ in Sect. 3 (still denoted by $\mathcal{L}%
^{P}$, by abuse of language) admitting a \textit{partial classical semantics}
for radical formulas. This semantics can be particularized to fit in with
various different interpretations of quantum mechanics. If one adopts a
suitable orthodox interpretation or a modal interpretation, our notion of
truth weakens the classical notion but does not conflict with it. If one
accepts the generalization and reinterpretation of quantum mechanics (%
\textit{extended semantic realism, }or \textit{ESR},\textit{\ model})
proposed by us together with some collaborators [Garola, 2015; Garola and
Sozzo, 2009, 2010, 2011a, 2011b, 2011c, 2012; Garola and Persano, 2014;
Garola \textit{et al.}, 2014, 2015], our notion of truth coincides with the
classical notion. Based on this generalization of $\mathcal{L}^{P}$, we
select in Sect. 4 a sublanguage $\mathcal{L}_{Q}^{P}$\ of $\mathcal{L}^{P}$\
in which the notion of proof is specified to be the notion of empirical
proof in quantum mechanics. We then show in Sect. 5 that this specification
induces a homomorphism of $\mathcal{L}_{Q}^{P}$\ onto QL: hence, an
interpretation of QL as a structure formalizing the properties of the
metalinguistic notion of justification according to quantum mechanics, not
the properties of a notion of truth alternative to the classical notion.%
\footnote{%
This procedure does not strictly match the procedure adopted \ in [Dalla
Pozza and Garola, 1995] to recover intuitionistic logic within $\mathcal{L}%
^{P}$. Indeed the axioms of AIPC (which make AIPC isomorphic to IPC) are
sentences of $\mathcal{L}^{P}$\ that are \textit{pragmatically valid} (%
\textit{p-valid}) in $\mathcal{L}^{P}$, for they characterize a notion of
logical proof. Some of the axioms that characterize the quantum notion of
proof are instead sentences of $\mathcal{L}_{Q}^{P}$ that are not p-valid.
We therefore avoid a purely axiomatic approach, which would make our task
uselessly complicate.} \ This result holds for each of the interpretations
of QM mentioned above (orthodox, modal and ESR model), hence we can claim
that our aim has been reached.

We conclude our work in Sect. 6 by comparing our result with a similar
result obtained by ourselves together with another author\ [Garola and
Sozzo, 2013], showing the advantages of the approach proposed in this paper.

To close, we note that our generalization of $\mathcal{L}^{P}$ is important
also independently of quantum mechanics, because it makes\ $\mathcal{L}^{P}$
a more powerful tool for coping with a variety of problems.

\section{On the notion of truth in quantum mechanics}

Busch \textit{et al}. \ [1991, 1996] distinguish two basic classes of
interpretations of quantum mechanics, the class of \textit{statistical} and
the class of \textit{realistic} interpretations. According to the
statistical interpretations quantum mechanics deals only with probabilities
of measurement outcomes, and no reference to single items of physical
systems (briefly, \textit{individual objects} in the following) is allowed.
According to the realistic interpretations quantum mechanics deals with
individual objects and their physical properties.

The statistical interpretations imply an instrumentalist view that has been
severely criticized from an epistemological viewpoint. For instance, Timpson
[2006] writes:

\begin{quote}
``The point is, instrumentalism is not a particularly attractive or
interesting interpretive option in quantum mechanics, amounting more to a
refusal to ask questions than to take quantum mechanics seriously. It is
scarcely the epistemologically enlightened position that older generations
of physicists, suffering from positivistic hang-overs, would have us
believe.''
\end{quote}

We add that nowadays experimental physicists often claim that they can deal
with individual objects, not only with statistical ensembles.

The class of the realistic interpretations, on the other side, is very
broad. In fact, the requirement that quantum mechanics deals with individual
objects characterizes a very weak form of realism, which does not imply any
wave or particle model for individual objects nor necessarily entails
ontological committments (one could indeed interpret the term ``individual
object'' as ``activation of a preparing procedure'' [Ludwig, 1983]). Hence
this class includes both standard interpretations and
reinterpretations/modifications of quantum mechanics that are realistic in a
stronger sense (as Bohm's theory, many worlds \ interpretation, GRW theory,
etc.). To avoid misunderstandings we therefore call the interpretations of
this class \textit{individual} rather than \textit{realistic}
interpretations in the following.

Whenever an individual interpretation is adopted, however, the measurement
problem arises, which was known since the birth of quantum mechanic and is
clearly formalized by some famous ``no-go'' theorems, as
Bell-Kochen-Specker's\ [Bell, 1966; Kochen and Specker, 1967)] and Bell's
[Bell, 1964], stating the \textit{contextuality} and the \textit{nonlocality}
(i.e., contextuality at a distance) of quantum mechanics, respectively (see, 
\textit{e.g.}, [Mermin, 1993]). In short, the result of a measurement of a
physical property on an individual object in a quantum state \footnote{%
We recall that ``quantum states'' and ``physical properties'' can be
considered as theoretical terms of the language of quantum mechanics which
can be operationally interpreted as follows. A quantum state $S^{Q}$
corresponds to a subset of physically equivalent \textit{preparation
procedures}\ belonging to a set of preparation procedures associated with $%
\Omega $. A physical property $E$ corresponds to a subset of physically
equivalent \textit{dichotomic registering devices} belonging to a set of
dichotomic registering devices associated with $\Omega $. Each device $r$ of
the latter subset, if activated in succession with the activation of a
preparation procedure $p$ in the subset corresponding to $S^{Q}$, performs a
measurement of $E$ on an item of $\ \Omega $ prepared by $p$, that is, on an 
\textit{individual object }$a$\textit{\ in (the quantum state) }$S^{Q}$,
after which $a$ either displays $E$\ or not. In this sense we say that $E$
is \textit{testable} [Beltrametti and Cassinelli, 1981; Ludwig, 1983; Garola
and Sozzo 2014] (we remind, however, that there are physical properties in
quantum mechanics that are \textit{incompatible}, in the sense that they
cannot be tested conjointly). We add that quantum states are usually divided
into two disjoint classes, the class of \textit{pure} quantum states and the
class of \textit{mixed}\ quantum states, or \textit{mixtures}.} is not
prefixed according to quantum mechanics, but it depends on the set of
(compatible) measurements that are simultaneously performed on the
individual object (even at a distance, in the case of measurements performed
on an individual object that is an item of a composite system whose
component parts are far away). Of course, this typical feature of quantum
mechanics does not follow because of \ flaws or errors in the measurement
devices (the measurements are supposed to be \textit{exact}). Hence,
generally, no truth value can be assigned to sentences attributing physical
properties to individual objects in a given state without taking into
account the set of measurements that are performed. Thus, truth and
verification by means of measurements could not be separated in quantum
mechanics, whose language would require the adoption of a (non-classical)
verificationist theory of truth.

The feature of quantum mechanics described above raises a deep problem, as
we have anticipated in Sect. 1. It implies in fact that the general scheme
provided by the original formulation of $\mathcal{L}^{P}$\ does not fit in
with an individual interpretation of quantum mechanics, because no truth
assignment could be defined on the set of radical formulas of $\mathcal{L}%
^{P}$ (whenever these formulas are interpreted as sentences of the language
of quantum mechanics) independently of the justification value of the
corresponding assertive formulas. However, we maintain that this problem can
be avoided by introducing a suitable generalization of $\mathcal{L}^{P}$.
Indeed, for every quantum state $S^{Q}$ of a physical system there are
observables whose values can be predicted with probability 1,\ independently
of the measurement context (to be precise, all observables that admit $S^{Q}$
as an eigenstate). Hence one can associate with $S^{Q}$ a subset of \textit{%
objective} properties, that is, a subset of physical properties that are
possessed with probability 1 or 0 by every individual object $a$ whose state
is $S^{Q}$, independently of any measurement. One can then assign a truth
value that does not depend on the measurement context to each sentence
attributing one of these physical properties to $a$: truth value \textit{true%
} if the probability is 1, truth value \textit{false} if the probabiliity is
0. Thus one obtains a noncontextual \textit{quantum partial truth assignment}
associated with $S^{Q}$. Hence truth and verification can be distinguished
if one restricts to the subset of objective properties.

Bearing in mind the above argument, we will generalize $\mathcal{L}^{P}$\ by
admitting truth assignments on radical formulas that are not defined
everywhere (the \textit{partial classical semantics} mentioned in Sect. 1).
This generalization fits in well with those standard interpretations of
quantum mechanics that consider \textit{real} (or \textit{actual}) in the
quantum state $S^{Q}$ every physical property $E$ such that the\ probability 
$p(S^{Q},E)$ associated by quantum mechanics to the pair $(S^{Q},E)$ is 1
[Jauch, 1968; Piron, 1976; Busch \textit{et al}., 1991, 1996; Aerts, 1999].
We call these interpretations \textit{standard realistic interpretations} in
the following, because their proposers usually maintain that they express a
realistic philosophical position.

Our generalization of $\mathcal{L}^{P}$ fits in well also with other
interpretations of quantum mechanics that introduce new sets of states
besides quantum states. We refer in particular to the \textit{modal
interpretations} of quantum mechanics (see the bibliography in [Lombardi and
Dieks, 2014]). In these interpretations the \textit{dynamical states}
correspond to the quantum states. But a further set of \textit{value states}
is introduced, and the set of all sentences\ attributing a physical property
to an individual object $a$ that have a truth value is determined by the
value state of $a$. This truth assigment is consistent with the quantum
partial truth assignment introduced above but it is defined on a broader set
of sentences. However, every truth assigment associated with a value state
is still partial, and can be seen as an instantiation of the generalized
semantics introduced in $\mathcal{L}^{P}$.

Finally, our generalization of $\mathcal{L}^{P}$ is compatible also with the
ESR model proposed by ourselves together with some collaborators in the
papers quoted in Sect. 1. This model in fact generalizes quantum mechanics
and circumvents the ``no-go'' theorems by reinterpreting quantum
probabilities as conditional on detection rather than absolute.
Contextuality and nonlocality are thus avoided. Hence, truth values are
defined for every quantum state and for all sentences of the language of
quantum mechanics according to classical rules and independently of the
measurement context. This truth assigment is a borderline case of the
quantum partial truth assignment introduced above because it is defined
everywhere. Hence it can be considered as an instantiation of the semantics
provided in the original version of $\mathcal{L}^{P}$, no generalization
being needed. It is interesting, however, to observe that one can construct 
\textit{hidden variables models} for the ESR model in which \textit{%
microscopic states} are introduced that play a role analogous to the role of
value states in the\ modal interpretations: the main difference is that the
truth assigments associated with the microscopic states are not partial but
defined everywhere [Garola et al., 2015].

\section{The generalized pragmatic language $\mathcal{L}^{P}$}

Let us summarize sintax, semantics and pragmatics of a pragmatic language $%
\mathcal{L}^{P}$ which generalizes the language denoted by the same symbol
in the paper by Dalla Pozza and Garola [1995].

\textit{Alphabet. }The alphabet $\mathcal{A}^{P}$ of $\mathcal{L}^{P}$
contains as \textit{descriptive signs }the propositional letters $p$, $q$, $%
r $,...; as\textit{\ logical-semantic signs} the connectives $\urcorner $, $%
\wedge $, $\vee $, $\rightarrow $ and $\leftrightarrow $; as \textit{\
logical-pragmatic signs} the assertion sign $\vdash $ and the connectives $N$%
, $K$, $A$, $C$ and\textit{\ }$E$; as\textit{\ auxiliary signs} the round
brackets $($,$)$.\textit{\ }

\textit{Radical formulas. }The set $\psi _{R}$ of all \textit{radical
formulas (rfs)} of $\mathcal{L}^{P}$ is made up by all formulas constructed
by means of descriptive and logical-semantic signs, following the standard
recursive rules of classical propositional logic. We denote by $\phi _{R}$
the subset of all rfs consisting of a propositional letter only (\textit{%
atomic formulas}).

\textit{Assertive formulas. }The set $\psi _{A}$ of all \textit{assertive
formulas (afs)} of $\mathcal{L}^{P}$ is made up by all rfs preceded by the
assertive sign $\vdash $ (\textit{elementary} afs), plus all formulas
constructed by using elementary afs and following standard recursive rules
in which $N$, $K$, $A$, $C$ and\ $E$ take the place of $\urcorner $, $\wedge 
$, $\vee $, $\rightarrow $ and $\leftrightarrow $, respectively. We denote
by $\phi _{A}$ the subset of all elementary afs of $\psi _{A}$.

\textit{Semantic interpretation.} Let us introduce a family $\{\sigma
_{S}\}_{S\in \mathcal{S}}$, where $\mathcal{S}$ is a set of \textit{states}
which play the role of possible worlds in Kripkean semantics and $\sigma
_{S} $ is a \textit{\ }function which maps a subset $\phi _{RS}$ of $\phi
_{R}$\ (the \textit{domain} of $\sigma _{S}$) onto the set $\{1,0\}$ of%
\textit{\ truth values} ($1$ standing for \textit{true} and $0$ for \textit{%
false}). We assume that $\phi _{R}=\cup _{S\in \mathcal{S}}\phi _{RS}$, so
that, for every $\alpha \in \phi _{R}$, at least one state \textit{S} exists
such that $\alpha \in \phi _{RS}$. Then, for every $S\in \mathcal{S}$, let
us extend $\sigma _{S}$ to the set $\psi _{RS}\subseteq \psi _{R}$ of all
rfs which contain only atomic formulas that belong to $\phi _{RS}$,
following the standard truth rules of classical propositional logic. We call 
\textit{assignment function} this extension, denote it by $\sigma _{S}^{e}$,
and call \textit{semantic interpretation} of $\mathcal{L}^{P}$ the family $%
\{\sigma _{S}^{e}\}_{S\in \mathcal{S}}$. We stress that we do not assume
that $\phi _{RS}$ is a proper subset of $\phi _{R}$: if $\phi _{RS}=\phi
_{R} $, then $\psi _{RS}=\psi _{R}$ and $\sigma _{S}^{e}$ reduces to a
classical assignment function. In general, however, $\{\sigma
_{S}^{e}\}_{S\in \mathcal{S}}$ can be considered as a \textit{weakened
classical semantics} for $\mathcal{L}^{P}$.

\textit{Pragmatic interpretation. }Whenever a semantic interpretation $%
\{\sigma _{S}^{e}\}_{S\in \mathcal{S}}$ is given, a \textit{pragmatic
interpretation} of $\mathcal{L}^{P}$ is defined as a family $\{\pi
_{S}\}_{S\in \mathcal{S}}$, where $\pi _{S}$ is a \textit{pragmatic
evaluation function} which maps $\psi _{A}$ onto the set $\{J,U\}$ of 
\textit{\ justification values} ($J$ standing for \textit{justified} and $U$
for \textit{unjustified}). We assume that each $\pi _{S}$\ satisfies the
following \textit{\ justification rules} (where $\alpha $ and $\delta $\
play the role of metalinguistic variables), which refer to $\{\sigma
_{S}^{e}\}_{S\in \mathcal{S}}$ and are based on the informal properties of
the metalinguistic concept of proof in natural languages.

\smallskip

JR$_{1}$. \textit{Let }$S\in \mathcal{S}$ and $\alpha \in \psi _{R}$\textit{%
. Then, }$\pi _{S}(\vdash \alpha )=J$\textit{\ if }$\alpha \in \phi _{RS}$ 
\textit{and\ a proof exists that }$\sigma _{S}^{e}(\alpha )=1$\textit{, }$%
\pi _{S}(\vdash \alpha )=U$\textit{\ otherwise.}

\smallskip

JR$_{2}$.\textit{\ Let }$S\in \mathcal{S}$ and $\delta \in \psi _{A}$\textit{%
. Then, }$\pi _{S}(N\delta )=J$\textit{\ if a proof exists that it is
impossible to prove that }$\pi _{S}(\delta )=J$\textit{, }$\pi _{S}(N\delta
)=U$\textit{\ otherwise.}

\smallskip

JR$_{3}$.\textit{\ Let }$S\in \mathcal{S}$ and $\delta _{1}$\textit{, }$%
\delta _{2}\in \psi _{A}$\textit{. Then,}

\textit{(i) }$\pi _{S}(\delta _{1}K\delta _{2})=J$\textit{\ iff }$\pi
_{S}(\delta _{1})=J$\textit{\ and }$\pi _{S}(\delta _{2})=J$\textit{,}

\textit{(ii) }$\pi _{S}(\delta _{1}A\delta _{2})=J$\textit{\ iff }$\pi
_{S}(\delta _{1})=J$\textit{\ or }$\pi _{S}(\delta _{2})=J$\textit{,}

\textit{(iii) }$\pi _{S}(\delta _{1}C\delta _{2})=J$\textit{\ iff a proof
exists that }$\pi _{S}(\delta _{2})=J$\textit{\ whenever }$\pi _{S}(\delta
_{1})=J$\textit{,}

\textit{(iv) }$\pi _{S}(\delta _{1}E\delta _{2})=J$\textit{\ iff }$\pi
_{S}(\delta _{1}C\delta _{2})=J$\textit{\ and }$\pi _{S}(\delta _{2}C\delta
_{1})=J$\textit{.}

\smallskip

Let us add some terminology and comments on JR$_{1}$-JR$_{3}$.

First of all, for every $S\in \mathcal{S}$ and $\delta \in \psi _{A}$ we
briefly say in the following that $\delta $\ is \textit{justified} (\textit{%
unjustified}) in $S$ whenever $\pi _{S}(\delta )=J$ ($U$).

Secondly, let us recall that rules JR$_{2}$, JR$_{3}$ (iii) and JR$_{3}$
(iv) make reference to a notion of proof that belongs to a higher logical
level with respect to the notion of proof involved in rules JR$_{1}$, JR$%
_{3} $ (i) and JR$_{3}$ (ii) [Dalla Pozza and Garola, 1995]. To make this
point clear, let us concentrate on JR$_{2}$ (JR$_{3}$ (iii) and JR$_{3}$
(iv) will not be needed in the following) and let us consider an example.
Let $S\in \mathcal{S}$ and $\alpha \in \phi _{RS}$. Then, stating that $%
\vdash \alpha $ is unjustified means that we do not possess any proof of $%
\alpha $, but does \ not prohibit that a proof of $\alpha $ can be produced:
hence it does not imply that $N\vdash \alpha $ is justified. The af $N\vdash
\alpha $ is instead justified iff a proof exists that a proof of $\alpha $
cannot be produced: hence, in particular, if a proof exists that $\alpha $
is false. Thus, $\pi _{S}(N\vdash \alpha )=J$ implies $\pi _{S}(\vdash
\alpha )=U$, but the converse implication does not hold.

Thirdly, let us note that the following \textit{correctness criterion} holds
in $\mathcal{L}^{P}$ because of JR$_{1}$.

\smallskip

CC. \textit{Let }$S\in \mathcal{S}$\textit{\ and }$\alpha \in \psi _{RS}$%
\textit{. Then, }$\pi _{S}(\vdash \alpha )=J$\textit{\ implies }$\sigma
_{S}^{e}(\alpha )=1.$

\section{The quantum pragmatic language $\mathcal{L}_{Q}^{P}$}

As we have anticipated in Sect. 1, we aim to pick out in this section a
sublanguage of $\mathcal{L}^{P}$ and specify the notion of proof as
empirical proof in quantum mechanics. For the sake of clearness we will
proceed by steps.

\subsection{Alphabet and formation rules}

The \textit{quantum pragmatic language} $\mathcal{L}_{Q}^{P}$ is the
sublanguage of $\mathcal{L}^{P}$ defined by the following syntactic
restrictions.

\smallskip

R$_{1}$. \textit{The set }$\psi _{R}^{Q}$\ \textit{of all rfs of }$\mathcal{L%
}_{Q}^{P}$\textit{\ is the subset }$\phi _{R}=\cup _{S\in \mathcal{S}}\phi
_{RS}$ \textit{of atomic rfs of }$\mathcal{L}^{P}$\textit{.}\smallskip

R$_{2}$. \textit{The set }$\psi _{A}^{Q}$\textit{\ of all afs of }$\mathcal{L%
}_{Q}^{P}$\textit{\ is the set of all afs of }$\mathcal{L}^{P}$\textit{\ in
which only rfs in }$\psi _{R}^{Q}$\ \textit{and the logical-pragmatic signs }%
$\vdash $\textit{, }$N$\textit{\ and }$K$\textit{\ occur.}

\smallskip

Because of R$_{1}$ and R$_{2}$, the set $\psi _{A}^{Q}$ of afs of $\mathcal{L%
}_{Q}^{P}$ is made up by all formulas constructed by means of the following
recursive rules.\smallskip

(i) \textit{Let }$\alpha \in \psi _{R}^{Q}$\textit{. Then }$\vdash \alpha
\in \psi _{A}^{Q}$\textit{.}

(ii) \textit{Let }$\delta \in \psi _{A}^{Q}$\textit{. Then }$N\delta \in
\psi _{A}^{Q}$\textit{.}

(iii) \textit{Let }$\delta _{1}$\textit{, }$\delta _{2}\in \psi _{A}^{Q}$%
\textit{.\ Then, }$\delta _{1}K\delta _{2}\in \psi _{A}^{Q}$\textit{%
.\smallskip }

The restrictions expressed\ by R$_{1}$ and R$_{2}$ are obviously introduced
to make it possible to contrive an \textit{intended interpretation} of $%
\mathcal{L}_{Q}^{P}$ in terms of quantum physics. In order to justify R$_{1}$%
, R$_{2}$ and our further assumptions on $\mathcal{L}_{Q}^{P}$, let us
discuss this interpretation in some details.

\subsection{The intended interpretation of $\mathcal{L}_{Q}^{P}$.}

Let $\Omega $ be a physical system, characterized in quantum mechanics by a
set $\mathcal{E}$ of (first order) physical properties, let $\mathcal{U}$ be
a set of individual objects and let.$\mathcal{P}(\mathcal{U})$ be the power
set of$\mathcal{\ U}$. For every $E\in \mathcal{E}$ and $a\in \mathcal{U}$,
we write $E(a)$ to formalize the informal sentence ``the individual object $%
a $ has the physical property $E_{\alpha }$'' (to avoid proliferation of
symbols we do not distinguish here between a physical property and the first
order predicate expressing it).

The set $\mathcal{S}$ of states of $\mathcal{L}_{Q}^{P}$ can be interpreted
as the set of all \textit{quantum states} of some standard realistic
interpretation of quantum mechanics, or as the set of all \textit{value
states} of some modal interpretation, or with the set of all \textit{%
microscopic states} in the ESR model (Sect. 2). We do not specify one of
these interpretations at this stage, for we wish our treatment to be as
general as possible.

For every interpretation of $\mathcal{S}$ we introduce a mapping

\begin{center}
$ext:S\in \mathcal{S}\longrightarrow extS\in \mathcal{P}(\mathcal{U})$
\end{center}

such that $\left\{ extS\right\} _{S\in \mathcal{S}}$ is a partition of $%
\mathcal{U}$, and say that $extS$ is the \textit{extension }of\textit{\ }$S$
(of course, $extS$\ depends on the interpretation of $S$\ that has been
chosen).\ Furthermore, for every $a\in \mathcal{U}$ we say that ``$a$ is in
the state $S$'' whenever $a\in extS$. Then we provide a physical
interpretation of the rfs of $\mathcal{L}_{Q}^{P}$ by assuming that a
bijective mapping exists

\begin{center}
$I:\alpha \in \psi _{R}^{Q}\longrightarrow E_{\alpha }\in \mathcal{E}$
\end{center}

such that the following semantic condition holds.\smallskip

SC. \textit{Let} $\alpha \in \psi _{R}^{Q}$, $S\in \mathcal{S}$\textit{, and
let }$a\in extS$. \textit{Then, }$\alpha \in \psi _{RS}^{Q}=\phi _{RS}$%
\textit{\ and }$\sigma _{S}(\alpha )=1$\textit{\ }$(0)$\textit{\ if and only
if the sentence }$E_{\alpha }(a)$\textit{\ is }true\textit{\ }(false) 
\textit{according to the interpretation of }$S$\textit{\ that is adopted.}

The semantic condition SC introduces different semantic interpretations of $%
\mathcal{L}_{Q}^{P}$, depending on the interpretation of S\ that has been
chosen. In any case SC implies that $E_{\alpha }(a)$ takes the same truth
value for every individual object in the state $S$.

The intended interpretation of the afs of $\mathcal{L}_{Q}^{P}$ is now
immediate if the term \textit{proof} in JR$_{1}$-JR$_{3}$ is meant as 
\textit{empirical proof}, that is, a\ proof following from quantum
mechanics, which is the physical theory that is considered in this paper.

Bearing in mind the intended interpretation above and footnote 2, it is
apparent that R$_{1}$ in Sect. 4.1 is introduced to select only rfs that
have a truth value for some state $S$ and can be interpreted as \textit{%
testable}, or \textit{verifiable}, sentences, i.e., sentences such that
physical procedures exist which test their truth value (which does not
always occur, because of incompatibility of properties, in the case of
nonatomic, or \textit{molecular}, rfs; note that a similar restriction has
been introduced by Dalla Pozza and Garola [1995] when recovering
intuitionistic propositional logic within $\mathcal{L}^{P}$). R$_{2}$ is
introduced instead for the sake of simplicity, since only the pragmatic
connectives $N$ and $K$ are relevant to our goals in this paper.

\subsection{The semantics of $\mathcal{L}_{Q}^{P}$}

The semantics of $\mathcal{L}_{Q}^{P}$ is obtained by restricting the
assignment functions defined on $\psi _{R}$ to $\psi _{R}^{Q}=\phi _{R}$.
Hence, for every $S\in \mathcal{S}$, the assignment function $\sigma
_{S}^{e} $ reduces to $\sigma _{S}$ and its domain $\psi _{RS}^{Q}\subseteq
\psi _{R}^{Q}$\textit{\ }coincides with $\phi _{RS}$. Moreover, the semantic
condition SC requires us to add the following semantic principle.\smallskip

SP. \textit{Every assignment function }$\sigma _{S}$\textit{\ defined on }$%
\psi _{R}^{Q}$\textit{\ must preserve the truth values and the relations
among truth values of rfs of }$\mathcal{L}_{Q}^{P}$\textit{\ established by
the laws of\ quantum mechanics via} \textit{the intended interpretation }of $%
\mathcal{L}_{Q}^{P}$\textit{.\smallskip }

To illustrate SP let us supply an example.

Let $\alpha _{1},\alpha _{2}\in \psi _{R}^{Q}$, $S\in \mathcal{S}$, and for
every $a\in extS$ let the laws of quantum mechanics imply that $E_{\alpha
_{2}}(a)$ is true whenever $E_{\alpha _{1}}(a)$ is true. Then, $\sigma _{S}$
must be such that, if $\alpha _{1}\in \psi _{RS}^{Q}$ and $\sigma
_{S}(\alpha _{1})=1$, then $\alpha _{2}\in \psi _{RS}^{Q}$ and $\sigma
_{S}(\alpha _{2})=1$.

The semantic principle SP does not provide, however, any explicit rule for
establishing whether a rf $\alpha \in \psi _{R}^{Q}$ has a truth value in a
given state $S$, and whether this value is \textit{true} or \textit{false}.
To make SP more explicit, let us take into account the possible
interpretations of the set $S$ mentioned in Sect. 4.2.

Let us firstly interpret $S$ on the set of all \textit{pure} quantum states%
\footnote{%
See footnote 2. We consider here and in the following only pure quantum
states to avoid the more complicated formalism required to deal with
mixtures.} and let us discuss in some details the quantum partial truth
assigment introduced in Sect. 2. Let us consider a state $S\in \mathcal{S}$
and a sentence of the form $E(a)$, with $E\in \mathcal{E}$ and $a\in extS$.
As we have seen in Sect. 2, one assigns a truth value to $E(a)$ which is
independent of the measurement context if $E$ is objective in the quantum
state $S$, that is, if the probability $p(E,S)$ associated by quantum
mechanics to the pair $(E,S)$\ is $1$ (truth value \textit{true}) or $0$
(truth value \textit{false}). Therefore, let us consider the set $\mathcal{E}%
_{S}=\mathcal{E}_{S}^{T}\cup \mathcal{E}_{S}^{F}$, with $\mathcal{E}%
_{S}^{T}=\left\{ E\mid E\in \mathcal{E},\text{ }p(E,S)=1\right\} $ and $%
\mathcal{E}_{S}^{F}=\left\{ E\mid E\in \mathcal{E},\text{ }p(E,S)=0\right\} $%
. Bearing in mind the mapping $I$ introduced in Sect. 4.2 and condition SC,
we obtain that, for every $\alpha \in \psi _{R}^{Q}$, $\alpha \in \psi
_{RS}^{Q}$ iff $I(\alpha )\in \mathcal{E}_{S}$, and $\ \sigma _{S}(\alpha
)=1 $ ($\sigma _{S}(\alpha )=0$) iff $I(\alpha )\in \mathcal{E}_{S}^{T}$ ($%
I(\alpha )\in \mathcal{E}_{S}^{F}$). Otherwise, $\sigma _{S}(\alpha )$ is
not defined, that is, $\alpha $ does not belong to $\psi _{RS}^{Q}$. This
specification of the way in which truth values are assigned can be restated
in terms of set-theoretical conditions on the assignment functions of the
family $\{\sigma _{S}\}_{S\in \mathcal{S}}$. To this end, let us consider
the set $\mathcal{E}$\ of all physical properties of a physical system $%
\Omega $. It is well known that in quantum mechanics $\mathcal{E}$\ is the
support of a lattice structure $\mathcal{L}(\mathcal{E})=(\mathcal{E},^{\bot
},\Cap ,\Cup )$, usually called \textit{standard (sharp) quantum logic}. In
this logic $^{\bot }$, $\Cap $ and $\Cup $ are considered as quantum logical
connectives. The symbol $^{\bot }$ denotes an involutory unary operation on $%
\mathcal{E}$\ called \textit{orthocomplementation}. The symbols $\Cap $ and $%
\Cup $ denote join and meet, respectively, in $\mathcal{L}(\mathcal{E})$ 
\footnote{%
We introduce here the symbols $\Cap $ and $\Cup $ in place of the symbols $%
\wedge $ and $\vee $\ that are usually introduced in quantum logic to denote
meet and join, respectively, in the lattice $\mathcal{L}(\mathcal{E})$. We
indeed want to avoid the (mis)interpretation of these symbols as classical
\textquotedblleft and\textquotedblright\ and \textquotedblleft
or\textquotedblright , respectively.}. This lattice is orthomodular but not
distributive (it also has some further mathematical properties that do not
interest us here) [Beltrametti and Cassinelli, 1981]. Moreover, quantum
mechanics associates a subset $\mathcal{S}_{E}\subseteq \mathcal{S}$ of
states with every $E\in \mathcal{E}$ in such a way that the set $\mathcal{S}%
_{\mathcal{E}}=\left\{ \mathcal{S}_{E}\mid E\in \mathcal{E}\right\} $,
partially ordered by the set inclusion $\subseteq $, is a lattice $\mathcal{L%
}(\mathcal{S}_{\mathcal{E}})=(\mathcal{S}_{\mathcal{E}},\subseteq )=(%
\mathcal{S}_{\mathcal{E}},^{\bot },\Cap ,\Cup )$ isomorphic to $\mathcal{L}(%
\mathcal{E})$, and \ the following properties hold.\medskip

(i) For every $E\in \mathcal{E}$,

$\mathcal{S}_{E^{^{\bot }}}=\mathcal{S}_{E}^{\bot }$.\medskip

(ii) Let $\cap $\ and $\cup $\ denote set theoretical intersection and
union, respectively. Then for every $E,F\in \mathcal{E}$,

$\mathcal{S}_{E\Cap F}=\mathcal{S}_{E}\Cap \mathcal{S}_{F}=\mathcal{S}%
_{E}\cap \mathcal{S}_{F}$,

$\mathcal{S}_{E\Cup F}=\mathcal{S}_{E}\Cup \mathcal{S}_{F}\supseteq \mathcal{%
S}_{E}\cup \mathcal{S}_{F}$.\medskip

The truth assigments introduced above can now be restated in set-theoretical
terms by referring to the lattice $\mathcal{L}(\mathcal{S}_{\mathcal{E}})$.
Indeed it can be shown in quantum mechanics that, for\ every $E\in \mathcal{E%
}$ and $S\in \mathcal{S}$, $E\in \mathcal{E}_{S}^{T}$ ($\mathcal{E}_{S}^{F}$%
) if and only if $S\in \mathcal{S}_{E}$ ($\mathcal{S}_{E}^{\bot }$) [Garola
and Sozzo, 2013]. Hence, for every $a\in extS$, a truth value of $E(a)$ is
defined if and only if $S\in \mathcal{S}_{E}\cup \mathcal{S}_{E}^{\bot }$,
which is \textit{true} if and only if $S\in \mathcal{S}_{E}$,\ \textit{false}
if and only if $S\in \mathcal{S}_{E}^{\bot }$. This result can be
transformed into an explicit rule for any $\sigma _{S}\in \{\sigma
_{S}\}_{S\in \mathcal{S}}$, as follows.\medskip

TR.\textit{\ Let }$\alpha \in \psi _{R}^{Q}$ \textit{and }$S\in \mathcal{S}$%
\textit{. Then, }$\alpha \in \psi _{RS}^{Q}$\textit{\ if and only if }$S\in 
\mathcal{S}_{E_{\alpha }}\cup \mathcal{S}_{E_{\alpha }}^{\bot }$\textit{,
and }$\sigma _{S}(\alpha )=1$\textit{\ }$(0)$\textit{\ if and only if }$S\in 
\mathcal{S}_{E_{\alpha }}$\textit{\ (}$S\in \mathcal{S}_{E_{\alpha }}^{\bot
} $\textit{).\medskip }

The truth rule TR provides a set-theoretical semantic interpretation $%
\{\sigma _{S}\}_{S\in \mathcal{S}}$ of $\mathcal{L}_{Q}^{P}$ that follows
from the general principle SP whenever S is interpreted as the set of all
quantum states of some standard realistic interpretation of quantum
mechanics.

Let us now interpret $\mathcal{S}$ as the set of all value states in a modal
interpretation of quantum mechanics (Sect. 2) and let us consider the
semantic\ interpretation $\{\sigma _{S}\}_{S\in \mathcal{S}}$ in this case.
Let $a\in \mathcal{U}$\ and let us put $\mathcal{E}(a)\mathcal{=}\left\{
E(a)\mid E\in \mathcal{E}\right\} $. Then, $a$\ belongs to the extension of
some value state and the subset of all sentences of $\mathcal{E}(a)$\ that
have a truth value \ is generally broader than the set of all sentences of $%
\mathcal{E}(a)$\ that have a truth value according to the semantic
interpretation discussed above. Moreover, whenever truth values of $E(a)\in 
\mathcal{E}(a)$\ are assigned by both interpretations, they coincide. But it
must be stressed that quantum mechanics can predict the truth value of $E(a)$
if and only if $E(a)$\ has a truth value according to the interpretation of $%
\mathcal{S}$ as a set of quantum states (which correspond to the \textit{%
dynamical states} in the modal interpretation of quantum mechanics).

Finally, let us interpret $\mathcal{S}$ as the set of all microscopic states
in the ESR model (Sect. 2). In this case the semantic interpretation $%
\{\sigma _{S}\}_{S\in \mathcal{S}}$\ is such that, for every $S\in \mathcal{S%
}$\textit{, }$\psi _{RS}=\psi _{R}$: hence it is a classical semantics for $%
\mathcal{L}_{Q}^{P}$. If one considers an individual object $a\in \mathcal{U}
$, then $a$\ belongs to the extension of some microscopic state and all
sentences of the set $\mathcal{E}(a)$\ defined above have a truth value.
Moreover, the truth value of a sentence $E(a)\in \mathcal{E}(a)$\ coincides
with the value of $E(a)$\ according to the interpretation of $\mathcal{S}$
as a set of quantum states (which correspond to the \textit{macroscopic
states} in the ESR model) whenever the latter value is assigned. But it must
be stressed that also in this case quantum mechanics predicts the truth
value of $E(a)$ if and only if $E(a)$\ has a truth value according to the
latter interpretation of $\mathcal{S}$.

\subsection{The pragmatics of $\mathcal{L}_{Q}^{P}$}

Proceeding as in Sect. 4.3, the pragmatics of $\mathcal{L}^{P}$ is obtained
by restricting each pragmatic evaluation functions $\pi _{S}$\ defined on $%
\psi _{A}$ to $\psi _{A}^{Q}$ (this restriction will still be denoted\ by $%
\pi _{S}$ to avoid proliferation of symbols). Moreover, in the case of
elementary afs of $\mathcal{L}_{Q}^{P}$ we specify the notion of
justification as empirical\ proof by introducing the following pragmatic
principle.\medskip

PP. \textit{Let }$S\in \mathcal{S}$\textit{\ and }$\alpha \in \psi _{R}^{Q}$%
\textit{. Then }$\pi _{S}(\vdash \alpha )=J$\textit{\ if }$\alpha \in \psi
_{RS}^{Q}$\textit{\ and the laws of quantum mechanics allow to prove, via
intended interpretation, that }$\sigma _{S}(\alpha )=1$\textit{, }$\pi
_{S}(\vdash \alpha )=U$ \textit{otherwise.}\medskip

The pragmatic principle PP entails that only the interpretation of $\mathcal{%
S}$ as a set of quantum states is relevant at a pragmatic level. Indeed, it
asserts that an elementary af $\vdash \alpha $ is justified if and only if
the sentence $E_{\alpha }(a)$\textit{\ }(with $a$ an individual object in
the state $S$), which corresponds to the atomic rf $\alpha $\ via the
mapping $I$ (Sect. 4.2), can be proven to have truth value \textit{true}.
But such a proof is possible only if the truth value of $E_{\alpha }(a)$ is
assigned by a quantum partial truth assigment, that is, by the truth
assigment associated with a quantum state (Sect. 2). Quantum mechanics
indeed can predict only these truth values (Sect. 4.3).

Bearing in mind the above remark, we restrict to the interpretation of $%
\mathcal{S}$ as a set of quantum states in the following, and no further\
mention of value states or microscopic states will be done.\footnote{%
The notion of proof specified by PP\ is empirical in the sense that a proof
requires the use of physical laws. However, it can be considered empirical
also in a different sense, because the same proof can be obtained by means
of measurements. It can be shown in fact that a quantum partial truth
assigment assigns a\ value \textit{true} (\textit{false}) to a sentence $%
E(a) $, with $a$ in the state $S$, if and only if one can perform a
measurement of $E$\ on $a$ without modifying $S$\ [Garola and Sozzo, 2004].}

The notion of justification as empirical proof then extends to afs of $\psi
_{A}^{Q}$\ that are not elementary via rules JR$_{2}$ and JR$_{3}$.

The pragmatic principle PP does not provide, however, any explicit rule for
establishing whether the elementary af $\vdash \alpha $ is justified or
unjustified in a given state $S$. To make PP more explicit, let us show that
the intended interpretation in Sect. 4.2 and PP\ imply that the pragmatics
of $\mathcal{L}_{Q}^{P}$\ can be expressed in set-theoretical terms. To this
end, and bearing in mind the symbols introduced in Sects. 4.2 and 4.3, we
restate PP as follows.\medskip

PP$^{\prime }$. \textit{Let }$S\in \mathcal{S}$\textit{\ and let }$\alpha
\in \psi _{R}^{Q}$\textit{. Then, }$\pi _{S}(\vdash \alpha )=J$\textit{\ if }%
$S\in $\textit{\ }$\mathcal{S}_{E_{\alpha }}$\textit{, }$\pi _{S}(\vdash
\alpha )=U$\textit{\ if }$S\notin \mathcal{S}_{E_{\alpha }}$.\medskip

The justification rule PP$^{\prime }$ specifies $\pi _{S}$ on the set of all
elementary afs of $\mathcal{L}_{Q}^{P}$ and constitutes the starting point
for our task. In fact, we can now introduce a mapping\medskip

$f:\delta \in \psi _{A}^{Q}\longrightarrow \mathcal{S}_{\delta }\in \mathcal{%
L}(\mathcal{S}_{\mathcal{E}})$,\medskip

which associates a \textit{pragmatic extension }$\mathcal{S}_{\delta }$ with
every assertive formula $\delta \in \psi _{A}^{Q}$, defining $f$ by means of
the following recursive rules.\medskip

(i) \textit{Let }$\alpha \in \psi _{R}^{Q}$\textit{. Then,\ }$f(\vdash
\alpha )=\mathcal{S}_{\vdash \alpha }=\mathcal{S}_{E_{\alpha }}$\textit{.}

(ii) \textit{For every }$\delta $\textit{\ }$\in \psi _{A}^{Q}$\textit{, }$%
f(N\delta )=\mathcal{S}_{N\delta }=\mathcal{S}_{\delta }{}^{\bot }$\textit{.}

(iii)\textit{\ For every }$\delta _{1}$\textit{, }$\delta _{2}\in \psi
_{A}^{Q}$\textit{, }$f(\delta _{1}K\delta _{2})=\mathcal{S}_{\delta
_{1}K\delta _{2}}=\mathcal{S}_{\delta _{1}}\Cap \mathcal{S}_{\delta _{2}}$%
\textit{.\medskip }

The pragmatic evaluation function $\pi _{S}$ can then be calculated by
assuming the following recursive justification rules.\medskip

JR$_{1}^{Q}$. \textit{Let }$S\in \mathcal{S}$\textit{\ and }$\alpha $\textit{%
\ }$\in \psi _{R}^{Q}$.\textit{\ Then, }$\pi _{S}(\vdash \alpha )=J$\textit{%
\ if }$S\in \mathcal{S}_{\vdash \alpha }$\textit{, }$\pi _{S}(\vdash \alpha
)=U$\textit{\ if }$S\notin \mathcal{S}_{\vdash \alpha }$\textit{.}

JR$_{2}^{Q}$. \textit{Let }$S\in \mathcal{S}$\textit{\ and }$\delta $\textit{%
\ }$\in \psi _{A}^{Q}$\textit{. Then, }$\pi _{S}(N\delta )=J$\textit{\ if }$%
S\in \mathcal{S}_{N\delta },$\textit{\ }$\pi _{S}(N\delta )=U$\textit{\ if }$%
S\notin \mathcal{S}_{N\delta }$\textit{.}

JR$_{3}^{Q}$. \textit{Let }$S\in \mathcal{S}$\textit{\ and }$\delta _{1}$%
\textit{, }$\delta _{2}\in \psi _{A}^{Q}$\textit{. Then, }$\pi _{S}(\delta
_{1}K\delta _{2})=J$\textit{\ if }$S\in \mathcal{S}_{\delta _{1}K\delta
_{2}} $\textit{, }$\pi _{S}(\delta _{1}K\delta _{2})=U$\textit{\ if }$%
S\notin \mathcal{S}_{\delta _{1}K\delta _{2}}$.\textit{\medskip }

Rules JR$_{1}^{Q}$-JR$_{3}^{Q}$ specialize JR$_{1}$-JR$_{3}$, respectively,
in our present context. They are suggested by the following arguments.

Rule JR$_{1}^{Q}$. From PP$^{\prime }$and (i).

Rule JR$_{2}^{Q}$. Let $\alpha $\textit{\ }$\in \psi _{RS}^{Q}$. Then we
must consider three alternatives, that is, $S\in \mathcal{S}_{\vdash \alpha
} $, $S\in \mathcal{S}_{\vdash \alpha }^{\bot }$ and $S\notin \mathcal{S}%
_{\vdash \alpha }\cup \mathcal{S}_{\vdash \alpha }^{\bot }$. If $S\in 
\mathcal{S}_{\vdash \alpha }$, hence $S\notin \mathcal{S}_{\vdash \alpha
}^{\bot }=\mathcal{S}_{N(\vdash \alpha )}$, then $\vdash \alpha $ is
justified in $S$ because of JR$_{1}^{Q}$: therefore no proof exists that $%
\vdash \alpha $ cannot be justified, which implies that $N(\vdash \alpha )$
is unjustified. If $S\in \mathcal{S}_{\vdash \alpha }^{\bot }=\mathcal{S}%
_{N(\vdash \alpha )}$, then $S\in \mathcal{S}_{E_{\alpha }}^{\bot }$ because
of (i), hence $\alpha $ is false ($\sigma _{S}(\alpha )=0$) because of the
TR rule (Sect. 4.3): it follows that a proof exists that $\vdash \alpha $
cannot be justified in $S$, which implies that $N(\vdash \alpha )$ is
justified in $S$. If $S\notin \mathcal{S}_{\vdash \alpha }\cup \mathcal{S}%
_{\vdash \alpha }^{\bot }$, then $\alpha $ has \ no truth value according to
the TR rule, that is, according to the quantum truth assignment.
Nevertheless a value can be \textit{actualized} by suitable measurements,
and it can be \textit{true}), or $\alpha $ can have a truth value according
to the modal interpretation of quantum mechanics and it has a truth value in
the ESR model. Hence no proof of $\alpha $ is supplied by quantum mechanics,
but no proof exists that $\vdash \alpha $ cannot be justified in $S$, which
implies that $N(\vdash \alpha )$ is unjustified in $S$.

Rule JR$_{3}^{Q}$. Let $\alpha _{1}$, $\alpha _{2}\in \psi _{RS}^{Q}$. Then
the af $(\vdash \alpha _{1})K(\vdash \alpha _{2})$ is justified if and only
if $\vdash \alpha _{1}$ and $\vdash \alpha _{2}$ are justified, that is, if
and only if $S\in \mathcal{S}_{\vdash \alpha _{1}}$ and $S\in \mathcal{S}%
_{\vdash \alpha _{2}}$: hence, if and only if $S\in \mathcal{S}_{\vdash
\alpha _{1}}\cap \mathcal{S}_{\vdash \alpha _{2}}=\mathcal{S}_{\vdash \alpha
_{1}}\Cap \mathcal{S}_{\vdash \alpha _{2}}=\mathcal{S}_{(\vdash \alpha
_{1})K(\vdash \alpha _{2})}$.

\section{The pragmatic interpretation of quantum logic}

Bearing in mind the definition of the mapping $f$ in Sect. 4.4, the
justification rules JR$_{1}^{Q}$-JR$_{3}^{Q}$ can be unified by the
following rule.\medskip

JR$^{Q}$. \textit{Let }$S\in \mathcal{S}$\textit{\ and }$\delta $\textit{\ }$%
\in \psi _{A}^{Q}$\textit{. Then, }$\pi _{S}(\delta )=J$\textit{\ if }$S\in 
\mathcal{S}_{\delta },$\textit{\ }$\pi _{S}(\delta )=U$\textit{\ if }$%
S\notin \mathcal{S}_{\delta }$\textit{.\medskip }

Based on JR$^{Q}$, one can introduce a preorder (binary, transitive)
relation $\prec $ on $\psi _{A}^{Q}$\textit{.\medskip }

OR. \textit{Let }$\delta _{1}$\textit{, }$\delta _{2}\in \psi _{A}^{Q}$%
\textit{. Then, }$\delta _{1}\mathit{\prec }\delta _{2}$\textit{\ if and
only if, for every }$S\in \mathcal{S}$\textit{, }$\pi _{S}(\delta _{1})=J$%
\textit{\ implies }$\pi _{S}(\delta _{2})=J$\textit{.\medskip }

It follows from JR$^{Q}$ and OR that $\delta _{1}\mathit{\prec }\delta _{2}$%
\textit{\ }if and only if $\mathcal{S}_{\delta _{1}}\subseteq \mathcal{S}%
_{\delta _{2}}$. It is then apparent that the mapping $f$ is an order
homomorphism of $(\psi _{A}^{Q},\prec )$ onto $(\mathcal{S}_{\mathcal{E}%
},\subseteq )$. Since $f(N\delta )=\mathcal{S}_{\delta }{}^{\bot }$\textit{\ 
}and $f(\delta _{1}K\delta _{2})=\mathcal{S}_{\delta _{1}}\Cap \mathcal{S}%
_{\delta _{2}}$\textit{,} we briefly say that $f$ makes the connectives $N$
and $K$ correspond to the lattice operations $^{\bot }$ and $\Cap $,
respectively. Furthermore, let us introduce a derived connective $A^{Q}$ in $%
\mathcal{L}_{Q}^{P}$, defined by the equation

$\delta _{1}A^{Q}\delta _{2}=N((N\delta _{1})K(N\delta _{2}))$.

Then we get

$f(\delta _{1}A^{Q}\delta _{2})=(\mathcal{S}_{(N\delta _{1})K(N\delta
_{2})})^{\bot }=(\mathcal{S}_{\delta _{1}}^{^{\bot }}\Cap \mathcal{S}%
_{\delta _{2}}^{^{\bot }})^{\bot }=\mathcal{S}_{\delta _{1}}\Cup \mathcal{S}%
_{\delta _{1}}$.

Hence, $f$ makes the connective $A^{Q}$ correspond to the lattice operation $%
\Cup $. Thus, our homomorphism shows that the quantum logical connectives $%
^{\bot }$, $\Cap $ and $\Cup $ can bear an interpretation as
logical-pragmatic signs rather than logical-semantic signs.

The above interpretation can be made more cogent by introducing an
equivalence relation $\approx $ on $\psi _{A}^{Q}$, defined as
follows.\medskip

ER. \textit{Let }$\delta _{1}$\textit{, }$\delta _{2}\in \psi _{A}^{Q}$%
\textit{. Then, }$\delta _{1}\mathit{\approx }\delta _{2}$\textit{\ if and
only if }$\delta _{1}\mathit{\prec }\delta _{2}$\textit{\ and }$\delta _{2}%
\mathit{\prec }\delta _{1}$\ \textit{(equivalently, for every }$S\in 
\mathcal{S}$\textit{, }$\pi _{S}(\delta _{1})=J$\textit{\ if and only if }$%
\pi _{S}(\delta _{2})=J$\textit{).\medskip }

JR$^{Q}$ and ER imply indeed that $\delta _{1}\mathit{\approx }\delta _{2}$%
\textit{\ }if and only if $\mathcal{S}_{\delta _{1}}=\mathcal{S}_{\delta
_{2}}$. Let us consider the quotient set $\psi _{A}^{Q\prime }=\psi
_{A}^{Q}/\approx $ and the relation $\prec ^{\prime }$ canonically induced
on $\psi _{A}^{Q\prime }$ by the relation $\prec $ defined \ on $\psi
_{A}^{Q}$. Then, $\prec ^{\prime }$ is a partial order (binary, transitive,
antisymmetric and reflexive) on $\psi _{A}^{Q\prime }$. Moreover, the
mapping $f$ induces an order isomorphism of $(\psi _{A}^{Q\prime },\prec
^{\prime })$ onto $(\mathcal{S}_{\mathcal{E}},\subseteq )$. Hence $(\psi
_{A}^{Q\prime },\prec ^{\prime })$ is an (orthomodular) lattice, in which
lattice operations $N^{\prime }$, $K^{\prime }$ and $A^{Q\prime }$ are
defined which correspond to the operations $^{\bot }$, $\Cap $ and $\Cup $
defined on $\mathcal{S}_{\mathcal{E}}$, respectively. These operations are
related to the connectives $N$, $K$ and $A^{Q}$ by the following
equations.\medskip

$N^{\prime }[\delta ]_{\approx }=[N\delta ]_{\approx }$,

$[\delta _{1}]_{\approx }K^{\prime }[\delta _{1}]_{\approx }=[\delta
_{1}K\delta _{1}]_{\approx }$,

$[\delta _{1}]_{\approx }A^{Q\prime }[\delta _{1}]_{\approx }=[\delta
_{1}A^{Q}\delta _{1}]_{\approx }$.\medskip

We have thus obtained a pragmatic structure $(\psi _{A}^{Q\prime },\prec
^{\prime })=$ $(\psi _{A}^{Q\prime },N^{\prime },K^{\prime },A^{Q\prime })$
which is isomorphic to the quantum logic $(\mathcal{S}_{\mathcal{E}%
},\subseteq )=(\mathcal{S}_{\mathcal{E}},^{\bot },\Cap ,\Cup )$ introduced
in Sect. 4.3. These two structures can then be identified.

\section{Conclusions}

The results stated at the end of the preceding section are philosophically
important. They imply that QL can be considered as a pragmatic structure,
formalizing the properties of empirical justification in quantum mechanics
rather than a notion of quantum truth specific of this theory. This
conclusion agrees with the general integrationist perspective (global
pluralism) mentioned in Sect. 1, according to which nonstandard logics are
interpreted as theories of metalinguistic notions different from truth, thus
avoiding incompatibility with classical notions and preserving the globality
of logic. Our aims in Sect. 1 are thus reached. Of course this achievement
has a price. Indeed, if one adopts a standard realistic or a modal
interpretation of quantum mechanics, one can reconcile QL with a classical
notion of truth at the expense of weakening this notion by introducing
partial truth assignments. A complete reconciliation of QL with classical
logic is possible only by accepting the reinterpretation of quantum
probabilities introduced \ by the ESR model. In any case, our results
provide a further example of the explanatory power of the (generalized)
pragmatic extension $\mathcal{L}^{P}$ of classical logic, in which different
logical systems may coexist without conflicting because they are interpreted
as formalizing different metalinguistic concepts.

It remains to observe that conclusions similar to ours have been drawn in
[Garola, 1992, 2008] and [Garola and Sozzo, 2013]. In particular, in the
last of these papers a classical predicate calculus $L(x)$ is constructed
and enriched by introducing a physical preorder (which is implied by the
logical order but generally does not coincide with it) induced by the\
theory-dependent notion of ``certainly true in a state $S$''. A structure
isomorphic to QL is then recovered by selecting a subset of sentences of $%
L(x)$ that are verifiable according to quantum mechanics and adding a
physically justified orthocomplementation.\footnote{%
It is noteworthy that similar procedures also lead to recover classical
Boolean structures whenever the language of classical mechanics is
considered, which implies that the logical and the physical order coincide
and all sentences are assumed to be verifiable. Moreover structures
isomorphic to QL can be obtained in this classical case if the notion of
verification is suitably restricted, so that only some sentences turn out to
be verifiable. This result proves that QL occurs because of the notion of
verification that is adopted and does not characterize quantum mechanics,
consistently with a \ known position of some scholars concerned with the
foundations of quantum mechanics [Aerts, 1988, 1991, 1995, 1998, 1999].} Our
present approach, however, is more general in several senses. Firstly, it is
constructed in such a way to allow an orthodox physical interpretation of
the truth values that are assigned to the radical formulas of $\mathcal{L}%
^{P}$. On the contrary, the approach in [Garola and Sozzo, 2013] introduces
a classical semantics on $L(x)$ which has a physical meaning only if one
adopts the reinterpretation and generalization of quantum mechanics
propounded by the ESR model. Secondly, the procedures in [Garola and Sozzo,
2013] are, paraphrasing Salmon's classification of scientific explanations
[Salmon, 1989], ``bottom-up'', because they explain QL in terms of basic\
logical and physical structures. On the contrary, our present procedures are
``up-down''. Indeed, they are based on the (generalized) language $\mathcal{L%
}^{P}$, which has a pragmatic interpretation that does not depend on
specific physical theories and is suitable for recovering different
non-standard logics by specifying different notions of proof. Hence our
interpretation of QL\ as a pragmatic structure constitutes an instantiation
of a general method in a special case (empirical quantum proof).

\bigskip

\begin{center}
\textbf{ACKNOWLEDGEMENT}
\end{center}

Most topics in this paper were discussed with Prof. Carlo Dalla Pozza, who
recently passed away. The author is greatly indebted with him for his
valuable suggestions, clearness and critical ability, and considers his
death as an irrecoverable loss.

\bigskip

\begin{center}
\textbf{BIBLIOGRAPHY}
\end{center}

Aerts,\textit{\ }D. (1988).\textit{\ }The physical origin of the EPR paradox
and how to violateBell inequalities by macroscopic systems. In Lahti, P., et
al. (Eds.), \textit{Symposium on the foundations of modern physics} (pp.
305-320). Singapore: World Scientific.

Aerts,\textit{\ }D. (1991). A macroscopic classical laboratory situation
with only macroscopic classical entities giving rise to a quantum mechanical
probability model. In Accardi, L. (Ed.), \textit{Quantum probability and
related topics} (pp. 75-85). Singapore: World Scientific.

Aerts,\textit{\ }D. (1995).\textit{\ }Quantum structures: An attempt to
explain their appearance in nature. \textit{International Journal of
Theoretical Physics, }34, 1165-1186.

Aerts,\textit{\ }D. (1998). The hidden measurement formalism: What can be
explained and where quantum paradoxes remain. \textit{International Journal
of Theoretical Physics, }37, 291-304.

Aerts,\textit{\ }D. (1999).\textit{\ }Quantum mechanics: Structures, axioms
and paradoxes. In Aerts, D. \& Pykacz, J. (Eds.), \textit{Quantum physics
and the nature of reality} (pp. 141-205). Dordrecht: Kluwer.

Bell, J. S. (1964). On the Einstein-Podolski-Rosen Paradox. \textit{Physics }%
1, 195-200.

Bell, J. S. (1966). On the Problem of Hidden Variables in Quantum Mechanics. 
\textit{Review of Modern Physics, }38, 447-452.

Bellin, G. (2014). Categorical proof theory of co-intuitionistic linear
logic. \textit{Logical Methods in Computer Science}, 10, 1--36.
www.lmcs-online.org.

Bellin, G. (2015). Assertions, hypotheses, conjectures, expectations:
rough-sets semantics and proof-theory. In Pereira, L. C., et al. (Eds), 
\textit{Advances in Natural Deduction. A Celebration of Dag Prawitzs Work,
Trends in Logic}, 39 (pp. 193--241). Springer Science+Business Media.

Bellin, G. and Biasi, C. (2004). Towards a logic for pragmatics. Assertions
and conjectures. \textit{Journal of Logic and Computation}, 14, 473--506.

Bellin, G., Carrara, M., Chiffi, D. (2015b) On an intuitionistic logic for
pragmatics. \textit{Journal of Logic and Computation}, DOI:
10.1093/logcom/exv036.

Bellin, G., Carrara, M., Chiffi, D. and Menti, A. (2015a). Pragmatic and
dialogic interpretations of bi- intuitionism. Part 1. \textit{Logic and
Logical Philosophy}, 23, pp. 449--480, 2014. Errata Corrige in \textit{Logic
and Logical Philosophy}, 2015. Online April 18, 2015.

Bellin,\textit{\ }G. \& Dalla Pozza, C. (2002). A pragmatic interpretation
of substructural logics. In Sieg, W., et al. (Eds.), \textit{Reflections on
the foundations of mathematics - Essays in honor of Solomon Feferman}
(pp.139-163). \textit{Association for Symbolic Logic, Lecture Notes in
Logic, }15.

Bellin,\textit{\ }G. \& Ranalter, K. (2003). A Kripke-style semantics for
the intuitionistic logic of pragmatics ILP. \textit{Journal of Logic and
Computation, }13 (5), 755-775.

Beltrametti, E. G. \& Cassinelli, G. (1981). \textit{The logic of quantum
mechanics}. Reading, MA: Addison.

Biasi, C. and Aschieri, F. (2008). A term assignment for polarized
bi-intuitionistic logic and its strong normalization. \textit{Fundamenta
Informaticae, Special issue on Logic for Pragmatics}, 84, 185--205.

Busch,\textit{\ }P., Lahti, P. J. \& Mittelstaedt, P. (1996). \textit{The
quantum theory of measurement}. Berlin: Springer.

Carnap, R. (1932). \"{U}berwindung der Metaphysik durch logische Analyse der
Sprache. \textit{Erkenntnis II}, 219-241.

Carnap, R. (1949). Truth and confirmation. In Feigl, H. \& Sellars, W.
(Eds.), \textit{Readings in philosophical analysis} (pp.119-127). New York:
Appleton-Century-Crofts, Inc.

Carrara, M. \& Chiffi, D. (2013). The knowability paradox in the light of a
logic for pragmatics. In Ciuni, R. et al. (Eds.), \textit{Advances in
philosophical logic} (pp. 33-48). \textit{Proceedings of Trends in Logic XI,
Studia Logica Library}. Berlin: Springer.

Dalla Chiara, M. L., Giuntini, R. \& Greechie, R. (2004). \textit{Reasoning
in quantum theory}. Dordrecht, Kluwer.

Dalla Pozza, C. \& Garola, C. (1995). A pragmatic interpretation of
intuitionistic propositional logic. \textit{Erkenntnis,} 43, 81-109.

Dummett, M. (1977). \textit{Elements of intuitionism}. Oxford: Clarendon
Press.

Dummett, M. (1978). \textit{Truth and other enigmas}. London: Duckworth.

Dummett, M. (1979). What does the appeal to use do for the theory of
meaning? In Margalit A. (Ed.), \textit{Meaning and use} (pp. 123-135).
Dordrecht: Reidel.

Dummett, M. (1980). Comments on professor's Prawitz paper. In Von Wright G.
H. (Ed.), \textit{Logic and philosophy/logique et philosophie} (pp. 11-18).
Dordrecht: Martinus Nijhoff Publishers.

Frege, G. (1879). \textit{Begriffschrift, eine der arithmetischen
nachgebieldete Formelsprache des reinen Denkens}. Halle: Nebert.

Frege, G. (1891). Funktion und Begriff. \textit{Vortrag, gehalten in der
Sitzung vom 9. Januar 1891 der Jenaischen Gesellschaft f\"{u}r Medizin und
Naturwissenschaft}. Jena: Hermann Pohle. (Collected in Angelelli I. (Ed.), 
\textit{Frege G.: \ Kleine Schriften} (pp. 124-142). Hildesheim: Olms, 1967).

Frege, G. (1893,). \textit{Grundgesetze der Arithmetik 1}. Jena: Pohle
(Hildesheim: Olms, 1967).

Frege, G. (1918). Der Gedanke. Eine Logische Untersuchung. In \textit{Beitr%
\"{a}ge zur Philosophie des deutschen Idealismus I} (1918-19), pp. 58-77.
(Collected in Angelelli I. (Ed.), \textit{Frege G.: \ Kleine Schriften} (pp.
342-362). Hildesheim: Olms, 1967).

Garola, C. (1992). Truth versus testability in quantum logic. \textit{%
Erkenntnis, }37, 197-222.

Garola, C. (2008). Physical propositions and quantum languages. \textit{%
International Journal of Theoretical Physics, }47, 90-103.

Garola, C. (2015). A survey of the ESR model for an objective
reinterpretation of quantum mechanics. \textit{International Journal of
Theoretical Physics}, DOI 10.1007/s10773-015-2618-y.

Garola, C. \& Persano, M. (2014). Embedding quantum mechanics into a broader
noncontextual theory. \textit{Foundations of Science, 19}, 217-239. DOI
10.1007/s10699-013-9341-z.

Garola, C., Persano, M., Pykacz, J. \& Sozzo, S. (2014). Finite local models
for the GHZ experiment. \textit{International Journal of Theoretical
Physics, }53, 622-644. DOI 10.1007/s10773-013-1851-5.

Garola, C. \& Sozzo, S. (2009). The ESR model: A proposal for a
noncontextual and local Hilbert space extension of QM. \textit{Europhysics
Letters, }86, 20009.

Garola, C. \& Sozzo, S. (2010). Embedding quantum mechanics into a broader
noncontextual theory: A conciliatory result. \textit{International Journal
of Theoretical Physics, }49, 3101-3117.

Garola, C. \& Sozzo, S. (2011a). Generalized observables, Bell's
inequalities and mixtures in the ESR model, \textit{Foundations of Physics, }%
41, 424-449.

Garola, C. \& Sozzo, S. (2011b). The modified Bell inequality and its
physical implication in the ESR model. \textit{International Journal of
Theoretical Physics, }50, 3787-3799.

Garola, C. \& Sozzo, S. (2011c). Representation and interpretation of
quantum mixtures in the ESR model. \textit{Theoretical and Mathematical
Physics, }168, 912-923.

Garola, C. \& Sozzo, S. (2012). Extended representation of observables and
states for a noncontextual representation of QM. \textit{Journal of Physics
A: Mathematical and Theoretical, }45, 075303 (13pp).

Garola, C. \& Sozzo, S. (2013). Recovering quantum logic within an extended
classical framework. \textit{Erkenntnis, }78, 399-419.

Garola, C., Sozzo, S. \& Wu, J. (2015). Outline of a generalization and a
reinterpretation of quantum mechanics recovering objectivity. \textit{%
ArXiv:1402.4394v3 [quant-ph]}.

Girard, J-Y. (1987). Linear logic. \textit{Theoretical Computer Science, }%
50, 1-102.

Haak, S. (1978). \textit{Philosophy of logic}. Cambridge: Cambridge
University Press.

Jammer, M. (1974). \textit{The Philosophy of Quantum Mechanics}. New York:
Wiley.

Jauch, J. M. (1968). \textit{Foundations of Quantum Mechanics}. London:
Addison-Wesley.

Kochen, S. and Specker, E. P. (1967). The problem of hidden variables in
quantum mechanics. \textit{Journal of Mathematical Mechanics,}17, 59-87.

Lombardi, O. and Dieks, D. (2014). Modal interpretations of quantum
mechanics. \textit{The Stanford Encyclopedia of Philosophy} (spring 2014
edition), Zalta E. N. (Ed.). URL=\TEXTsymbol{<}%
http://plato.stanford.edu/archives/spr2014/entries/qm-modal/\TEXTsymbol{>}.

Ludwig, G. (1983). \textit{Foundations of Quantum Mechanics I}. New York:
Springer.

Mermin, N. D. (1993). Hidden variables and the two theorems of John Bell. 
\textit{Reviews of Modern Physics, 65}, 803-815.

Piron, C. (1976). \textit{Foundations of Quantum Physics}.\textit{\ }%
Reading, MA: Benjamin.

Popper, K. (1969). \textit{Conjectures and refutations}. London: Routledge
and Kegan Paul.

Prawitz, D. (1977). Meaning and proof: on the conflict between classical and
intuitionistic logic. \textit{Theoria, }43, 1-40.

Prawitz, D. (1980). Intuitionistic logic: a philosophical challenge. In Von
Wright G. H. (Ed.), \textit{Logic and philosophy/Logique et philosophie}
(pp. 1-10). Dordrecht: Martinus Nijhoff Publishers.

Prawitz, D. (1987). Dummett on a theory of meaning and its impact on logic.
In Taylor B. M. (Ed), \textit{Michael Dummett} (pp. 117-165). Dordrecht:
Martinus Nijhoff Publishers.

Ranalter, K. (2008). A semantic analysis of a logic for pragmatics with
assertions, obligations and causal implication. \textit{Fundamanta
Informaticae, }84\textit{\ (3-4)}, 443-470

R\'{e}dei, M. (1998). \textit{Quantum Logic in Algebraic Approach}.
Dordrecht: Kluwer.

Reichenbach, H.: (1947). \textit{Elements of symbolic logic}. New York: The
Free Press.

Russell, B. (1940). \textit{An inquiry into meaning and truth}. London:
Allen \& Unwin.

Russell, B. (1950). Logical positivism. \textit{Revue Internationale de
Philosophie, }4, 3-19. (Collected in Russell, B. (1956). \textit{Logic and
knowledge}. London; Allen \& Unwin).

Salmon, W.C. (1989). Four decades of scientific explanation. In Kitcher, P.
\& Salmon, W. C. (Eds.), \textit{Scientific explanation. Minnesota studies
on the philosophy of science (13)} (pp. 3-219). Minnneapolis: University of
Minnesota Press.

Stenius, E. (1969). Mood and language-games. In Davis, J. W. et al.(Eds.), 
\textit{Philosophical logic} (pp. 251-271). Dordrecht: Reidel.

Tarski, A. (1933). Pojecie prawdy w jezykash nauk dedukcyjnych. \textit{Acta
Towarzystwei Naukowego i Literakiego Warszaswskiego}, 34, V-16; (1956. The
concept of truth in formalized languages. In Woodger J. M. (Ed.), \textit{%
Logic, semantics, metamathematics} (pp. 152-268). Oxford: Oxford University
Press (trans.).

Tarski, A. (1944).\ The semantic conception of truth and the foundations of
semantics. \textit{Philosophy and phenomenological research, }4, 341-375
(1952. In Linski, L. (Ed.), \textit{Semantics and the philosophy of language}
(pp. 13-47). Urbana: University of Illinois Press.

Timpson, C. (2008). Philosophical aspects of quantum information theory. In
Rickler, D. (Ed.), \textit{The Ashgate companion to the new philosophy of
physics} (pp. 197-261). Aldershot: Ashgate.

Troelstra, A. and Van Dalen, D. (1988). \textit{Constructivism in mathematics%
}. Amsterdam: North Holland.

Van Dalen, D. (1986). Intuitionistic logic. In Gabbay, D. \& Guenthner, F.
(Eds.), \textit{Handbook of philosophical logic III} (pp. 225-339).
Dordrecht: Reidel.

White, G. (2008). Davidson and Reiter on action. \textit{Fundamenta
Informaticae, }84 (2), 259-289.

\end{document}